\documentclass[sigconf, balance=false]{acmart}
\AtBeginDocument{%
  \providecommand\BibTeX{{%
    \normalfont B\kern-0.5em{\scshape i\kern-0.25em b}\kern-0.8em\TeX}}}

\copyrightyear{2021}
\acmYear{2021}
\setcopyright{acmcopyright}\acmConference[EuroPLoP'21]{European Conference on Pattern Languages of Programs}{July 7--11, 2021}{Graz, Austria}
\acmBooktitle{European Conference on Pattern Languages of Programs (EuroPLoP'21), July 7--11, 2021, Graz, Austria}
\acmPrice{15.00}
\acmDOI{10.1145/3489449.3490004}
\acmISBN{978-1-4503-8997-6/21/07}

\setcopyright{none}
\renewcommand\footnotetextcopyrightpermission[1]{}
\begin{document}

\title{Embedded Platform Patterns for Distributed and Secure Logging}


\author{Fikret Basic}
\orcid{0000-0003-4688-270X}
\affiliation{%
  \institution{Graz University of Technology}
  \city{Graz}
  \country{Austria}}
\email{basic@tugraz.at}

\author{Christian Steger}
\affiliation{%
  \institution{Graz University of Technology}
  \city{Graz}
  \country{Austria}}
\email{steger@tugraz.at}

\author{Robert Kofler}
\affiliation{%
  \institution{NXP Semiconductors Austria GmbH Co \& KG}
  \city{Gratkorn}
  \country{Austria}}
\email{robert.kofler@nxp.com}

\renewcommand{\shortauthors}{Fikret Basic, Christian Steger, and Robert Kofler}

\begin{abstract}
With the advent of modern embedded systems, logging as a process is becoming more and more prevalent for diagnostic and analytic services. Traditionally, storage and managing of the logged data are generally kept as a part of one entity together with the main logic components. In systems that implement network connections, this activity is usually handled over a remote device. However, enabling remote connection is still considered a limiting factor for many embedded devices due to the demanding production cost.
A significant challenge is presented to vendors who need to decide how the data will be extracted and handled for an embedded platform during the design concept phase. It is generally desirable that logging memory modules are able to be addressed as separate units. These devices need to be appropriately secured and verifiable on a different system since data compromise can lead to enormous privacy and even financial losses.
In this paper, we present two patterns. First, a pattern that allows flexible logging operation design in terms of module and interface responsibility separation. Second, a pattern for the design of secure logging processes during the utilization of constrained embedded devices.
The introduced patterns fulfil the following conditions: (i) flexibility – design is independent of the chip vendors making the logging memory modules easily replaceable, (ii) self-sufficiency – every logging controller is maintained as a separate entity in a decentralized topology, (iii) security – through providing authenticity, confidentiality, and integrity by means of using a dedicated security module. 

\end{abstract}

\begin{CCSXML}
<ccs2012>
   <concept>
       <concept_id>10010520.10010553.10010562</concept_id>
       <concept_desc>Computer systems organization~Embedded systems</concept_desc>
       <concept_significance>500</concept_significance>
       </concept>
   <concept>
       <concept_id>10010520.10010553.10010562.10010564</concept_id>
       <concept_desc>Computer systems organization~Embedded software</concept_desc>
       <concept_significance>300</concept_significance>
       </concept>
  <concept>
       <concept_id>10010520.10010553.10010560</concept_id>
       <concept_desc>Computer systems organization~System on a chip</concept_desc>
       <concept_significance>100</concept_significance>
       </concept>
   <concept>
       <concept_id>10002978.10003029.10011703</concept_id>
       <concept_desc>Security and privacy</concept_desc>
       <concept_significance>300</concept_significance>
       </concept>
 </ccs2012>
\end{CCSXML}

\ccsdesc[500]{Computer systems organization~Embedded systems}
\ccsdesc[300]{Computer systems organization~Embedded software}
\ccsdesc[100]{Computer systems organization~System on a chip}
\ccsdesc[300]{Security and privacy}

\keywords{logging, design pattern, system design, embedded, cybersecurity}%, data privacy}

\maketitle

% For the storage version
\pagestyle{plain}
\vspace{\fill}
\pagebreak

\section{Introduction}
Even today, many embedded devices are still considered constrained, offering only limited resources compared to some more complex platforms. The constraints are presented through the limited size of the provided internal memory (both volatile and non-volatile), limitations on the processing power, reduced employment of communication standards and ports, and even the lack of some extended features (e.g., restrictions on the security capabilities). As the systems become more complex, a necessity arises to capture important log data during its lifetime. The log data is usually used for control and diagnostic purposes, but it can also have secondary uses being a dataset for various machine learning algorithms. This integration is often found today in many modern applications, ranging from surveillance systems to smart grids and vehicles. In addition to having an implemented logging procedure, an essential requirement from many users to vendors is to have the data sufficiently protected so as not to be spied on or tampered with by malicious intruders. These design considerations are becoming increasingly important today, as the preservation of secured data and user’s privacy are becoming an increasing topic of interest. From manufacturers of these devices, a considerable effort is required to design a system that fulfils challenges of having (i) limited or absent network capabilities, (ii)  security as a co-process, (iii) synchronization between main logic, logging and security operations, (iv) options of porting and changing devices, and (v) option of removing and handling the logged data as a separate unit. 

A few of the design patterns previously published in the original Gang of Four (GoF) patterns book have already been in use for the logging process~\cite{10.5555/186897}. Historically, this was often achieved using the \textsc{Chain of Responsibility} pattern. Furthermore, the \textsc{Factory} pattern was often used with a combination of \textsc{Command} or \textsc{Memento} to handle the log messages. To supplement the security constraint, some more specialized design patterns like the \textsc{Secure Logger} were introduced as well~\cite{steel2005}. They are generally handled as implementation design patterns, and hence, they are not focused on explaining the integration in higher-level designs, especially those concerning modern embedded platforms. As we are going to discuss in the problem statement, this is often a special case. In fact, with the embedded platforms, the controller is often seen as an independent unit from other components, such as sensors, actuators, other controllers, central units, etc. Moreover, many modern patterns are primarily focused on Cloud solutions and do not take into account local and restricted devices.
To overcome these restrictions, we introduce: (i) \textsc{\textit{Embedded Platform to Memory}} (EP2M), and (ii) \textsc{\textit{Secure Embedded Logging}} (SEL) patterns. EP2M presents a solution during the design process to handle the division of modular tasks between individual units by proposing a methodology with which both decentralisation and a streamlined production design can be achieved. SEL provides directions for establishing a secure logging operation pipeline between an embedded controller device and a memory unit. When applied together, they offer an affordable solution for designing a secure logging operation on individual embedded target devices.

The proposed patterns are intended primarily for vendors during the device design and production cycle but also for users during the deployment phase. Vendors are commonly embedded device manufacturers, but they can also be service providers. Users are customers, i.e., the side that integrates the provided embedded devices into a new or an already established system. Both vendors and users benefit from the pattern solutions. The patterns provide a cost-efficient way to port and upgrade (using EP2M), and securely verify (using SEL) the logging memory modules, even after their initial installation.

\section{Embedded Platform to Memory Pattern}
\label{sec:ep2m}
\begin{figure*}[h]
  \centering
  \includegraphics[width=\linewidth]{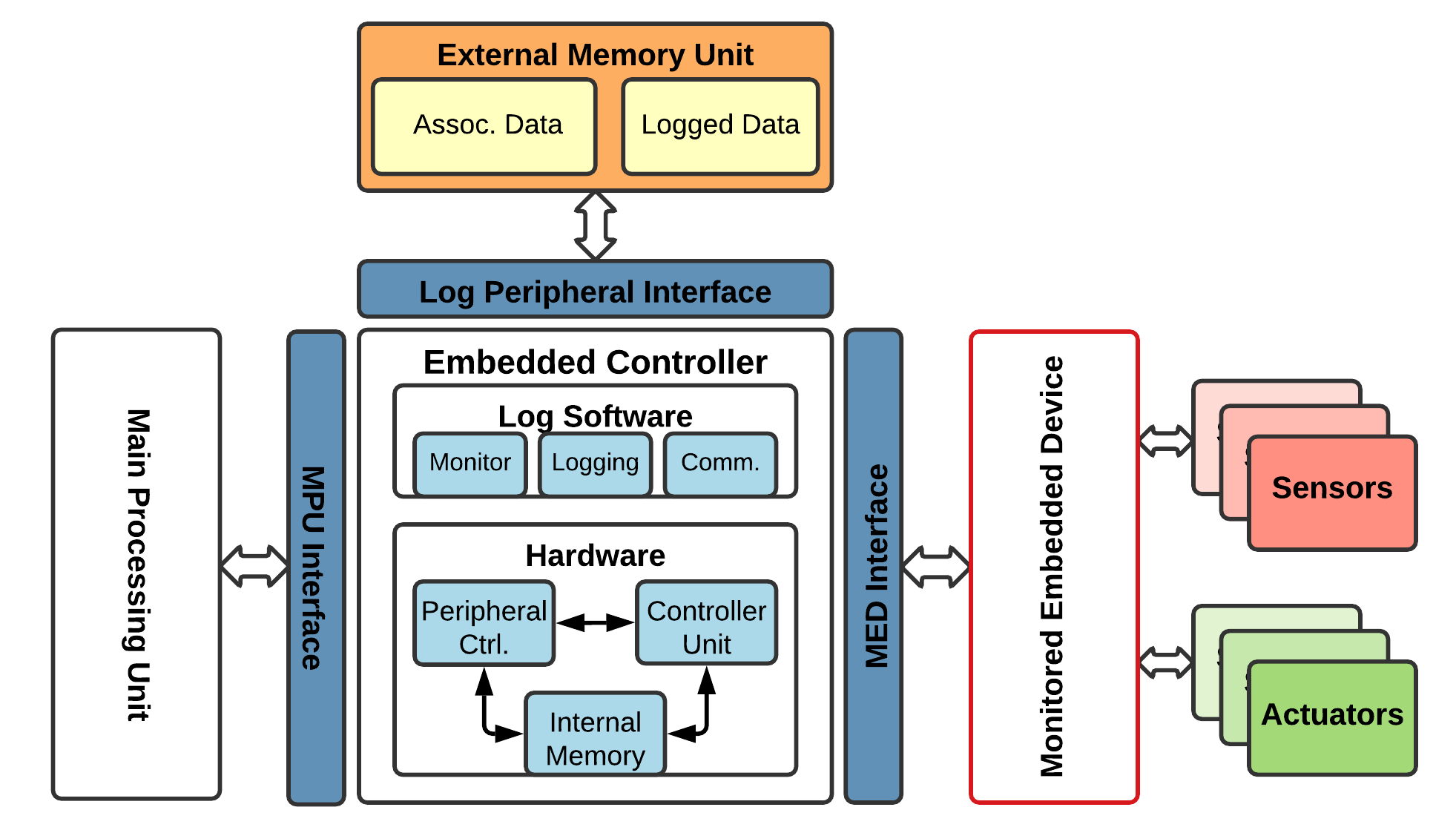}
  \caption{Block demonstration of the suggested modules and connection points during the design of the embedded logging process on a generic device using the proposed solution through the \textsc{Embedded Controller to Memory} pattern.}
  \Description{Embedded devices displayed like block diagrams. Contains the interface logic, peripheral blocks, memory blocks, and main software and hardware components. Embedded devices include: Embedded controller, External Memory Unit, Main Processing Unit, Monitored Embedded Device, and monitored Sensors and Actuators.}
  \label{fig:ec2m_general}
\end{figure*}

\subsection{Intent}
Adding data logging functionality to the constrained embedded platforms by module distribution and role specification in the early system design.

\subsection{Context}
You are developing a system that uses constrained embedded controllers conceptualized to handle processing and memory operations locally rather than using some external infrastructure (e.g., cloud). In this case, the logging process is considered an internally implemented function with a dedicated memory unit, communication channel, and processing logic on the controller side, handling the status of a monitored device. The stored data is further used for diagnostic purposes in case of safety or security issues or as historic data for analytic purposes. It might also need to be shipped together with the monitored embedded device when managing a replacement procedure during the system's lifetime.

\subsection{Motivating Example}
\label{sec:motivating_example}
To better understand the importance and use-case of the EC2M pattern, let us look at an example of an appliance in the automotive domain. Electric vehicles contain specialized embedded platforms called Battery Management System (BMS), dedicated for control and management of battery cells used to power up the engine and other components~\cite{10.1007/978-3-662-45286-8_4,    andrea2010battery}. Different derivations of BMS exist, with the modular and distributed BMS being more common than the others. Each Battery pack contains several dedicated sensors alongside battery cells~\cite{andrea2010battery}. The battery packs are controlled through Battery Cell Controllers (BCC), which are assigned to handle the immediate data control and throughput of these individual packs. 
A central BMS receives individual battery packs data from the BCCs. This data ranges from the sensor data (e.g., temperature data) to the voltage and current of a particular cell. They are used to extract information like state of charge (SoC) or state of health (SoH) \cite{8168251}. Based on the data received, BMS can also store and handle error events.

When a battery pack gets depleted, it needs to be replaced. The replaced battery pack can often still be used as an active component for some other appliances, e.g., power grids. Here, battery packs are aimed to be shipped together with their assigned BCCs. In case the BCCs are to remain as part of the vehicle and its BMS, a design compromise needs to be established to enable the logged operational data to be shipped with the battery pack as well.  
Since BMS would be mass-produced, a design needs to be made in the earlier phases of the development.

\subsection{Problem}
\textbf{Since embedded devices are difficult to upgrade after their initial instalment, which module responsibilities, interface connections, and architecture decisions would need to be made during the design phase to enable flexible and portable logging procedures?}

Often, embedded devices keep the processing and logging of the data on a local basis because of the performance constraints to keep the production cost at a minimum. This means that for logging functionality, an embedded device might have a dedicated non-volatile memory module pre-installed. The memory module would also have a pre-set task to log the recorded data from a monitored device which can be an Internet of Things (IoT) device, smart sensor, another embedded device, etc. When porting and changing of this device happens, it is generally challenging to also port its logged data, with the accumulated process and event data being kept closed as part of the system. This comes from the difficulty of not having an appropriately handled system architecture across all devices and also of the missing necessary port interface options on both the hardware and software levels. 

\subsubsection{Forces}
\label{pattern_1_forces}
\begin{itemize}
    \item [F1] \textit{Connectivity}: An embedded controller needs to be able to, through standard protocols and interfaces, easily access the dedicated log memory module.
    \item [F2] \textit{Decentralization}: There can be multiple monitored devices, with each being handled as a separate unit.
    \item [F3] \textit{Scalability}: The solution should correctly scale with each new device. The impact of the new devices should be kept at a minimum in relation to the overall system performance.
    \item [F4] \textit{Production cost}: Introduced cost that comes with the extra components and installations.
    \item [F5] \textit{Maintenance overhead}: Additional cost and time delays for changes and updates of the associated logging modules after the deployment phase.
    \item [F6] \textit{Operational performance}: Additional modules and design concepts also need to deal with the added performance impact. The focus is placed on the processing logic through queuing, ordering, timing of log records, as well as the size of the memory and computational resources. %Complexity of the data log portability needs to be taken into the consideration.
    \item [F7] \textit{Software coherence}: Allowing designers to adequately separate software development from the underlying hardware components as to allow for an easier update mechanism to individual sub-modules. 
    \item [F8] \textit{Security threats}: System should be able to answer to the common security pitfalls found when handling the logged data, i.e., guarding against data tampering and spying. 
    %These include protecting its confidentiality and integrity against spying and tempering from unauthorized entities, as well as guaranteeing system availability. 
\end{itemize}

\subsection{Solution}
\textbf{Divide the core logging component into multiple distributed individual embedded controllers, each containing interfaces to the central control module, monitored device, and an external memory unit while keeping process handlers open for relevant events.}

A careful interface separation and task assignment during the early system design is necessary to avoid cost increase. A designer needs to identify which core module components are required in the overall system design and what role they have. This role decoupling needs to be done for two reasons: (i) higher responsiveness to faults; in case the data source device gets damaged or corrupted, it is still possible to retrieve the prior logged data, (ii) flexible interchangeability; local device groups appear independent from each other. Therefore it is easier to replace individual components without the added complexity.

Based on the set Forces, as part of the solution, we consider the following devices:
\begin{itemize}
    \item \textit{Embedded Controller} (EC): The embedded device responsible for the logging and data processing between the monitored targeted device and the system's central unit. This device represents our main target of interest, where the design logic for the embedded logging process is placed. It can mean an expansion of an already existing standard device used for process control as part of an embedded platform. %Hence, only the extensions related to the logging activity will be considered under this solution.
    \item \textit{Monitored Embedded Device} (MED): The end-device of the system that is responsible for the data gathering and event action, i.e., it is the targeted device from which the log data is extracted. 
    \item \textit{External Memory Unit} (EMU): The module that stores the log data gathered from an MED, as well as the associated data (configuration, metadata). 
    \item \textit{Main Processing Unit} (MPU): The device tasked for the main system logic control, service providing, and connection to the external services and sub-systems. Since the ECs are designed to be constrained devices, usually found in a decentralized network, a more powerful device is needed to control all, or a group of, ECs in the system. In the system implementation, this device can be the same as one of the ECs as long as the resources offered correspond to the requirements presented by the overall system. %Due to these difficulties, an MPU is generally a more powerful and independent unit. 
\end{itemize}

The proposed design is shown in Figure ~\ref{fig:ec2m_general}. A Microcontroller Unit (MCU) can be used as the hardware control unit to construct an EC. It is used to handle, through software, the logging logic, MED process monitoring, and communication flow control, among other assigned operations. Another function of this controller unit is to handle the synchronization and sampling rate of the MED. These also include administering the commands from the MPU and controlling the internal operational states (e.g., active, idle, sleep). Optionally, an EC can also internally incorporate volatile and non-volatile memory, as it is indicated by the internal memory block. These parts can, however, highly influence the end-design cost and are recommended to be considered sparingly. Due to the limitation in functions, an adequate Application-Specific Integrated Circuit (ASIC) chip can also be provided instead of the more costly MCU.
An EC also provides separate interfaces for the communication with the MPU and the dedicated MED. These can either be wireless or wired, depending on the design constraints. Examples of wired interfaces would be Inter-integrated Circuit (I2C), Serial Peripheral Interface (SPI), Universal Asynchronous Receiver-Transmitter (UART). For the wireless communication interface we recommend Bluetooth Low Energy (BLE), low-frequency Radio-frequency Identification (RFID), and higher-frequency RFID, like Near-Field Communication (NFC), among other standards. While the wireless interfaces offer more applicability in their use-cases, it should be noted that they also require additional handling and construction cost for error corrections and cybersecurity preservation.

\subsubsection{Log Memory Interconnection}
The pattern establishes cheap, flexible, and extensible handling of a memory module dedicated to logging purposes. To this end, an interface is provided to the EC for the logging handling. Here, we propose the use of an \textit{External Memory Unit} (EMU) module. The EC needs to treat the added memory module as an external unit rather than a pre-embedded component that is part of the EC. This is done to achieve the portability and flexibility in adding and removing the memory that houses the logged data. It is recommended that the new EC device already has a pre-built interface port for communication with the extensible EMUs. These ports can use different communication standards. It is recommended to use a well-established and long-term lasting standard. Among others, these include I2C, SPI, and UART (serial). A wireless standard can also be used, although it is not recommended for this interface. A wireless interface would add an additional increase in cost and complexity, where it would also have less support when porting it among the vendors.

\begin{figure}[h]
  \centering
  \includegraphics[width=\linewidth]{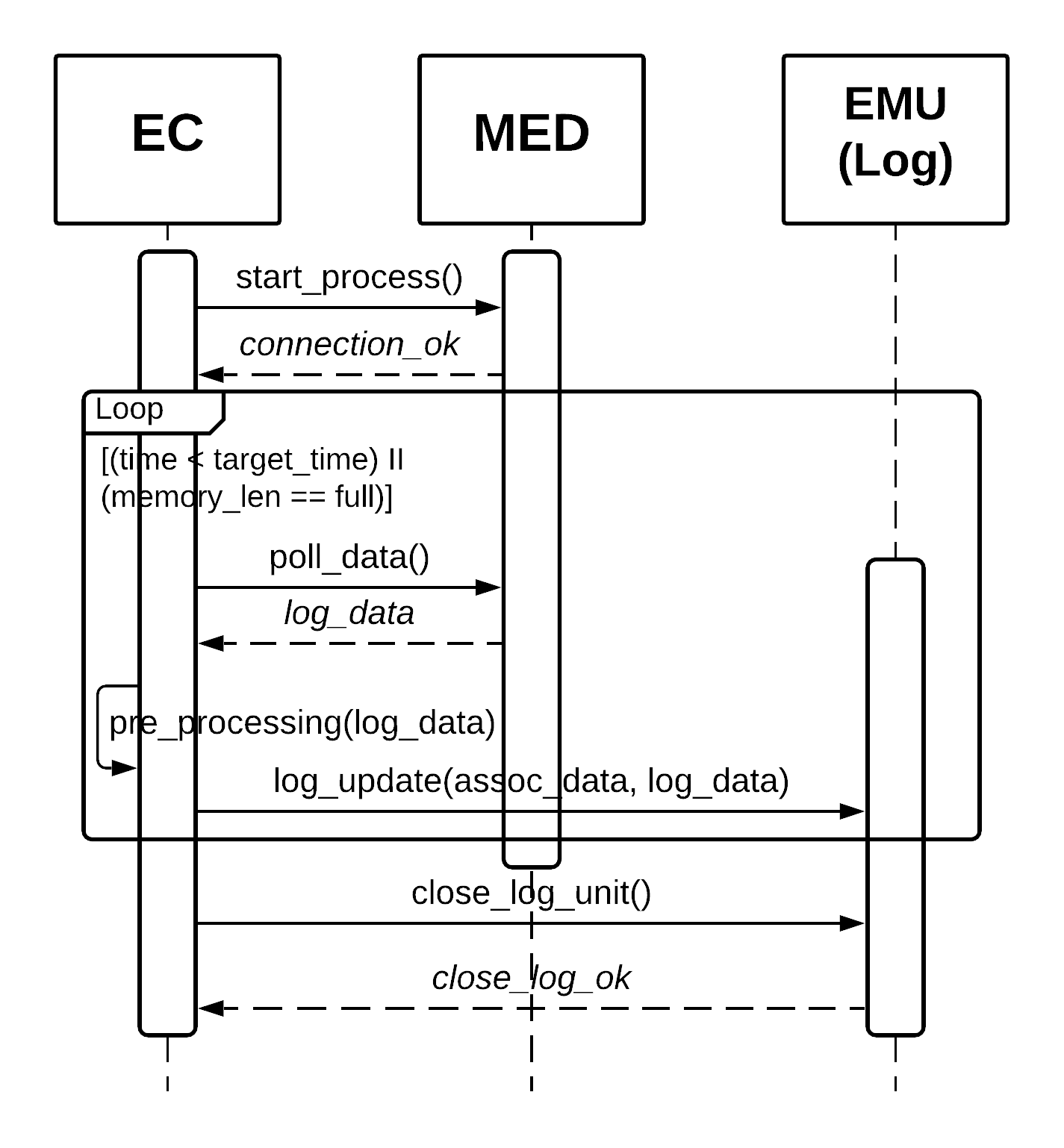}
  \caption{Sequence diagram with function calls for the starting, run, and closing phase during the logging operation.}
  \Description{Sequence diagram displaying the main sequence use-case for the logging operation between the monitoring embedded device (EC), monitored embedded device (MED), and the attached logging memory.}
  \label{fig:ec2m_seq_1}
\end{figure}

\subsubsection{Protocol Handling}
As already noted, the programming logic for the logging and memory handling should be appropriately handled through software implementation as part of the controller unit of the EC. The logging process should not interfere with the main controlling procedures but rather work as an extension. Timing delays are to be expected; hence the sampling rate for the logged data needs to be adapted accordingly. Figure~\ref{fig:ec2m_seq_1} illustrates the logic behind the logging process and individual components. The logging phase starts with first establishing the connection, followed by an interactive communication between the EC, MED and the Log Memory to catch and store the targeted data during the system's run-time. The last phase, closing of the communication, considers the remaining processes that are carried out after the main logging phase is over, e.g., calculating and storing associated operational data. The manner in which the logging processes are managed is an implementation task and is therefore left to the individual system designers. Here, we only indicate the main principle behind the logging procedure.
At the end of the dedicated lifetime of a MED in a system, it might be necessary to replace the component, and with that, to also port the old one together with the previously used EMU. This activity can be easily achieved since the system is intended to be modular, with each unit having the capability to be individually transported and replaced.

\subsection{Consequences}
To better assess the suggested pattern, we are going to list benefits and liabilities corresponding to the Forces from Section~\ref{pattern_1_forces}.\\
The benefits when using the EP2M pattern are:
\begin{itemize}
    \item [F2] Each pair of EC and MED is independent and unique, along with the dedicated memory component.
    \item [F3] The main logic control of the embedded platform is managed by a separate unit (MPU), that also controls which devices are added and handled, but does not cover the actual logging procedure. Therefore, it is possible to expand the system by adding additional ECs and MEDs, as long as the number adheres to the limitations set by the MPU.
    \item [F4] Since the solution proposes a modulated system, each component can easily be processed in a streamlined production line. The amount of the overall hardware and software necessary for the cross-platform support on the EC would result in its reduction as well.
    \item [F5] After the deployment, each memory module is easily replaceable. Also, the overall complexity is reduced when handling the logging procedure; it requires no special consideration, other than the design points already implemented in a pre-deployment phase.
    \item [F6] MPU, which represent the central logic, is free from the logging process. This frees up the resources necessary for the general system run.
    \item [F7] Through the careful hardware \& software design separation on the EC, the software is able to adequately access the necessary resources on the underlying hardware layer.
\end{itemize}

The liabilities when using the EP2M pattern are:
\begin{itemize}
    \item [F1] Since the production of the EC is handled separately from the memory module dedicated for data logging, an additional interface is needed for the EC for it to be able to communicate with an external log memory module. This adds additional cost and handling complexity. 
    \item [F3] Expansion of the system is limited by the system resources offered from the central MPU device. These are fairly predefined during the design phase.
    \item [F5] Maintenance and manual covering of individual devices could be an issue as the system scales. Additional devices would put a lot of constrains when handling them. %which might introduce the need of some additional design extensions. 
    \item [F8] The pattern helps in protecting system availability through its distributed solution. However, it is not focused on providing cybersecurity protection for secured log data.
\end{itemize}

\subsection{Known Uses}
The proposed solution can generally be found under two scenarios:
\begin{itemize}
    \item \textit{End-consumer aimed applications}: special home appliances, mobile phones, and surveillance systems~\cite{videoSurveillanceQumulo, videoSurveillanceSeagate}.
    \item \textit{Mission critical industrial applications}: process control systems, cellular base stations, medical systems, remote environmental data loggers and monitors~\cite{gyorok2017, gyorok2018}.  
\end{itemize}

Among these, the most common application today can be found as part of the more prominent industrial solutions where the utilization is necessary for traceable failure analysis. It is often used in the aeronautic and automotive domain inside the control ``Black Boxes``. Initially, these systems were aimed to provide a removable and safe memory module that logs the operational data during a dedicated session, where today they are also intended to provide sufficient security considerations \cite{blackBoxesPolicy}. 
Black Boxes are slowly becoming a norm in modern automobiles, designed to serve relevant operational data in case of accidents \cite{blackBoxesEU}.

\subsection{Realized Example}
\begin{figure}[h]
  \centering
  \includegraphics[width=\linewidth]{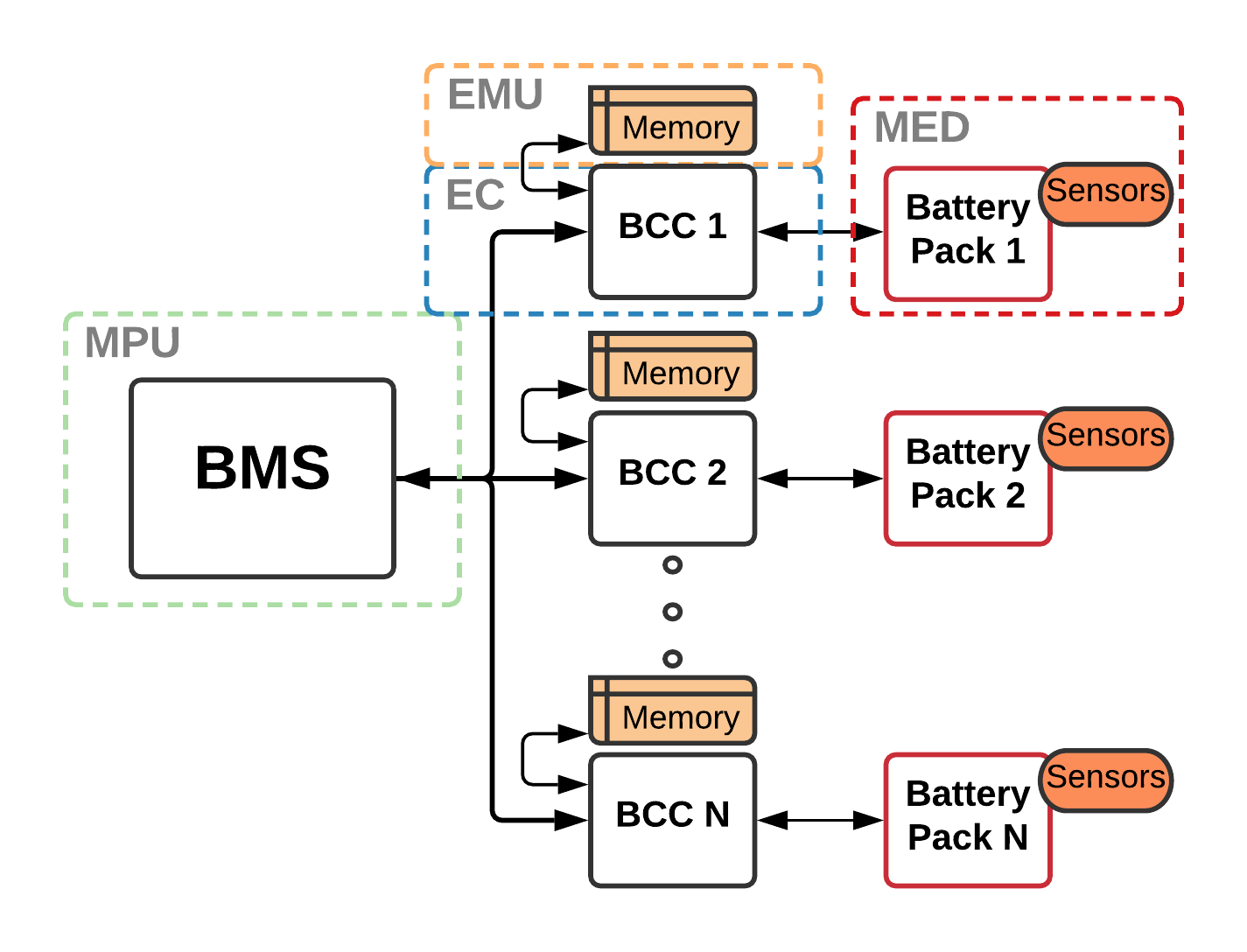}
  \caption{Realized motivating example using the \textsc{Embedded Controller to Memory} architectural pattern. Employed on a use-case concerning Battery Management System (BMS), having the ECs represented through Battery Cell Controllers (BCCs) that log the data from the Battery Packs.}
  \Description{Realized example containing the Battery Management System using the proposed solution. Elements are represented as block diagrams. Shows scalability by showcasing multiple BCC units and their connection to the Battery packs. Each battery pack contains an attached sensor. Each BCC contains an attached log memory unit.}
  \label{fig:ec2m_bms_example}
\end{figure}

To better understand where and how the EC2M pattern can be employed, we are going back to the example specified in Section~\ref{sec:motivating_example}. Here, we will apply our solution and analyze the outcome. As already noted, it is necessary to use a design solution to the BMS that covers the logging process for the sensor data received from battery packs. The outcome of the integrated design modules can be seen in Figure~\ref{fig:ec2m_bms_example}. As demonstrated, the BCC has been modified and extended with an interface for communication with an external memory module. Furthermore, the software in the MCU is developed to handle channel control to the memory module and appropriately cover the logging sample rate. 
The BCC communicates through additional interfaces with a battery pack on one side and the central BMS on the other end. In our applied solution, BMS is the MPU. BCC represents the EC, memory module is the EMU, with the battery pack being the MED. Each BCC and its assigned battery pack are handled as an individual group unit. An important aspect on why the solution had to be applied in the earlier stage of the development cycle, as suggested by the pattern, is the design of the essential interface connections. The system is expandable; hence additional BCCs and their battery packs can be attached in a daisy chain connection as indicated in Figure~\ref{fig:ec2m_bms_example}.  %An important aspect on why solution had to be applied in the earlier stage of the development cycle, as suggested by the pattern, is the design of the essential interface connections that allows the BCC as the EC to be connected with other modules, and these modules to also have available connections back and from the BCC.

\subsection{Related Patterns}
The \textsc{Chain of Responsibility}~\cite{10.5555/186897} is a behavioural design pattern that is structurally similar to the presented EC2M pattern. It can also be used to handle logging or auditing functionality. However, it does not account on its own for the modular responsibility distribution during the system's design phase. It is primarily implementation-oriented and can be applied on the software stack.

\section{Secure Embedded Logging Pattern}
\subsection{Intent}
Answering to the security needs by extending the data logging capabilities in the embedded platforms by adding security modules and services. The proposed pattern can be used to add the logging security features together with a design-focused logging solution. An example for the embedded logging design solution would be the \textsc{Embedded Platform to Memory} pattern described in Section~\ref{sec:ep2m}. 

\subsection{Context}
An embedded platform is being designed which uses local memory devices to handle the storage of lifetime logging data. For the reasons of the cost and memory size limitations, as well as not having, or having limited, access to a wide network, it is intended for the platform to rely on local solutions rather than remote services. Often, these types of systems are closed and protected under a specific group. It is critical that the stored data maintains its integrity and is only managed through authorized handlers in this environment. The embedded system would consist of a selection of hardware modules, interfaces, and implemented software functions. The hardware modules are divided by their respective tasks and placement. %, having a separate main controller, logging device, and the data source device.
These are usually tied to a specific architecture and their upgrade can be very difficult, or sometimes not even possible. 

\subsection{Motivating Example}
\label{sec:pattern_2_motivating_example}

For a complete example of the usage of the pattern, we will focus on the BMS use-case explained in Section~\ref{sec:motivating_example}. As stated, it is desirable to enable the porting of the stored memory units together with the battery packs as to be able to track the health of the used battery packs or for any additional data post-processing. It is of critical importance that only valid battery packs are being transported and that it is possible to authenticate the memory units used with the previous battery packs. This constraint is essential to make sure that no malicious attacks through a modified memory unit are possible. Additionally, the data that is stored needs also to be secured. The reason for making it secure is to guard it against any potential malicious attacks or even faults that can arise from oversights during service and maintenance.

The constraint is still present to handle these design steps in the initial development phase. This is done for the fact that the battery packs would be mass-produced. Any change that would otherwise be done later could jeopardize the security of the battery packs and add an additional cost.

\subsection{Problem}
\label{sec:pattern_2_problem}
% \textbf{How to maintain trust that the embedded platform logging data has not been compromised and that it comes from a trusted source when it is being ported?}
\textbf{How to design an embedded platform that is able to securely handle, but also port and verify, logging data from its source to a designated entity?}

In embedded platforms that use distributed module placement, logging process and porting of the logged data often introduce security risks. Porting would need to be done either by using a manual external device or having a connection to a network, both of which might be difficult, or even not feasible, under the platform constraints. Additionally, changes introduced to the system on a physical layer may hamper security during the transfer of the saved data and present a high level of porting complexity. Modules used would need to account for security functionality and have pre-defined elements that supplement them. These considerations result in making it a very challenging and expensive task.

Different malicious attacks can be mounted aimed directly at the content of the logged data during both the active logging period and during the offload transfer period. It is challenging to derive a definite list of threats, as these are usually use-case or application dependant. Here we focus primarily on generic threats that are found in embedded logging systems. Specifically, we consider the following main threats:
\begin{itemize}
    \item \textit{Spying on the targeted process}: If not properly secured, an attacker can derive information, and even knowledge, from the stored log data by a direct port access.
    \item \textit{Logged data tampering}: Unauthorized change of the current, or previously stored, log content. This includes active attacks on the communication points during the ongoing logging process, but also direct tamper attacks on the devices.
    \item \textit{Counterfeited sources}: Each logged data is tied to an affiliated monitored device that is also supplied from a certified manufacturer. During the offload transfer period of the device, i.e., when the change of the targeted monitored device happens, the device can be replaced with a counterfeited or a malicious one. A different attack would be by using the same device but replacing the data inside it. 
\end{itemize}

\subsubsection{Forces}
\label{pattern_2_forces}
\begin{itemize}
    \item [F1] \textit{Streamlined HW/SW integration}: Implementation of the hardware and software elements associated with the security functionality need to be easily replicated across multiple devices and vendors.
    \item [F2] \textit{Production cost}: Changes made to the hardware and software design of the embedded systems can result in an increased manufacturing cost. 
    \item [F3] \textit{Limited resources}: The embedded system needs to be able to execute all necessary functions under different constraints.
    \item [F4] \textit{Security - confidentiality and integrity}: Necessary measures need to be taken which should prevent the logged data to be tampered or spied on.
    \item [F5] \textit{Security - authenticity}: The logged data that is stored needs to be able to be properly identified and verified that it comes from a valid source entity. This authenticity is also necessary each time the data needs to be accessed during the active period, i.e., when the data is retrieved for the analytic or other operational purposes.
\end{itemize}

\subsection{Solution}
\textbf{Ensure that the monitored logged data will be securely protected through an integrated security module relaying data to the memory module and authenticated by using necessary hardware and software critical components embedded during the deployment phase.}

\begin{figure}[h]
  \centering
  \includegraphics[width=\linewidth]{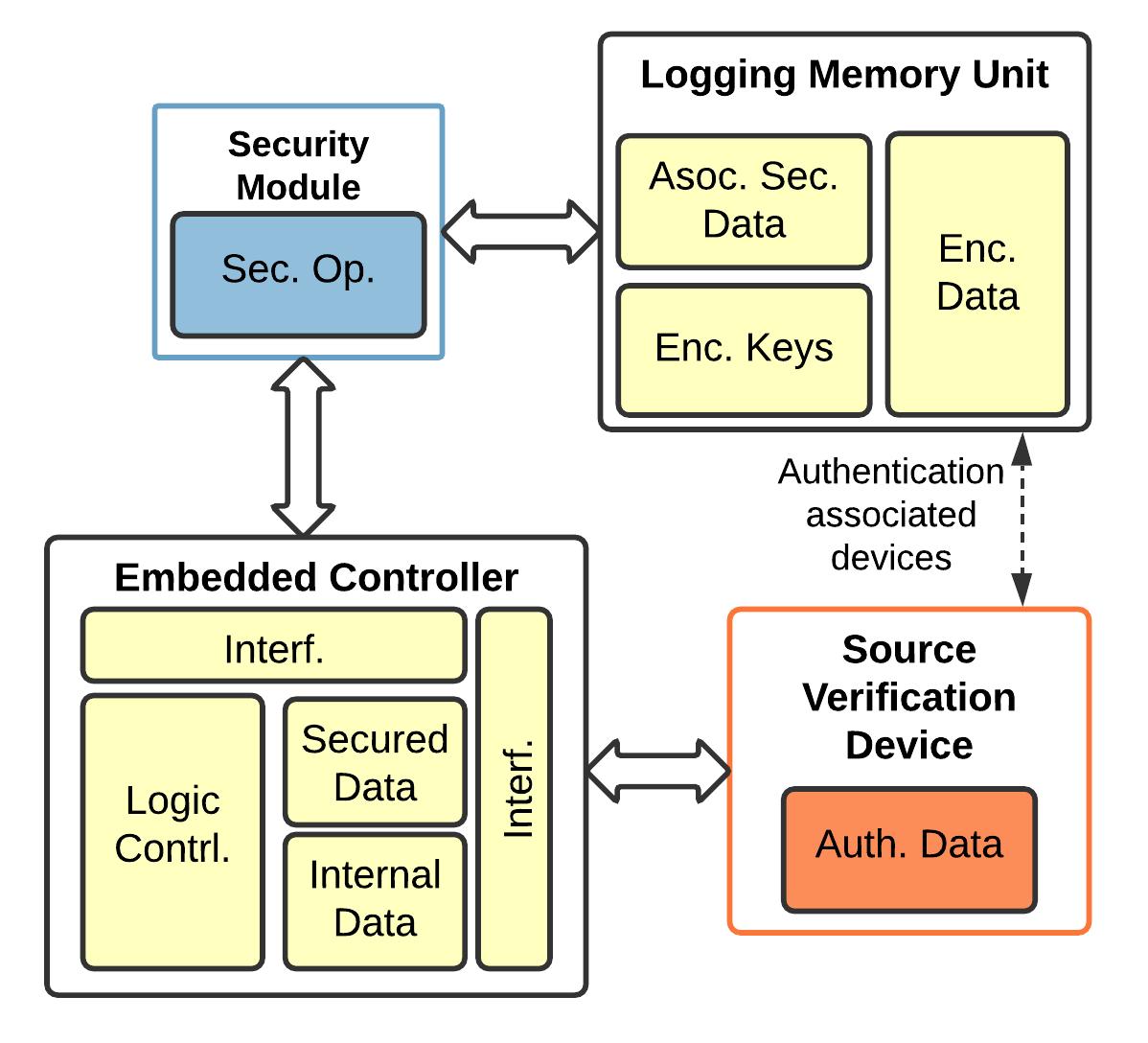}
  \caption{Design-based solution in task separation for handling security logging by providing secure operations and device authentication.}
  \Description{Block representation of the solution used for the SEL pattern showcasing the Embedded Controller, Security Module, Logging Memory Unit, and the Source Verification Device, alongside their sub-modules and connections.}
  \label{fig:sel_general}
\end{figure}

When implementing a logging procedure as part of the constrained embedded platform, the security requirement is achieved by integrating a Security Module (SM) as part of the EC. While adding the SM to individual EC devices adds to the overall cost, it does make the system more decentralized. Furthermore, this ensures that the security operations are distributed without heavily impacting the performance. EC device vendors could also not guarantee that the logged data would be secured since the EC itself would not handle that constraint. Therefore, it is necessary to also couple the security operations as part of the EC to appropriately address the security design and attest that the information stored will be protected.
Figure~\ref{fig:sel_general} depicts the design behind the solution and shows the recommended building blocks. The following components are listed:
\begin{itemize}
    \item \textit{Embedded Controller} (EC): Contains necessary interfaces for the communication, main driver logic, and the control bridge between the data that is to be stored and the security driver.
    \item \textit{Logging Memory Unit} (LMU): Dedicated device for storing the encrypted data; contains necessary description data, encrypted security keys, and the encrypted data.
    \item \textit{Security Module} (SM): Provides security operations; works as a security bridge between the EC and the LMU.
    \item \textit{Source Verification Device} (SVD): Device tied to a particular LMU and used for the authentication purpose; can contain necessary authentication data, i.e., private-public key pair, and/or a certificate. It is also generally seen as the device from which logging process data is retrieved (data source).
\end{itemize}

\textit{Software functions} and associated security data would be handled by the SM itself. At the same time, it would use the logic controller of the EC to drive the overall processing and data preparation when storing it as part of the logging. It is necessary to keep the SM cheap in design. As a minimum security requirement when storing the logging data, we propose to use encryption and authentication for the stored data. These can either be achieved by using separate security functions or applying a suite like Authenticated Encryption (AE) to handle this process. The data integrity check (additional AE data or a separate operation) would be saved in a separate memory block inside the LMU. These, however, do not need to be secured, but they do need to be checked by the EC from the SM each time a new LMU is authenticated. They also need to be periodically updated from the SM after new data is written.
The SM should also offer the functionality of storing and handling the key data used by the security operations. Additionally, an EC together with its SM could also provide a Key Derivation Function (KDF). The basic principle of deriving and delivering the keys between the parties is left to the designers. The keys are generally securely encrypted and stored in the LMU secure section. The authentication operations can either be managed using symmetric-based authentication, e.g., AES challenge/response mechanism or by using asymmetric authentication, e.g., Public Key Infrastructure (PKI).  
The security operations can be handled entirely through software or be hardware-derived, where the hardware operations usually offer better performance, e.g., hardware implementation of the Advanced Encryption Standard (AES). While we consider using the integrated security engine through a dedicated SM as the most cost-effective solution, other dedicated hardware security components can also be examined. These include Secure Elements (SE) and Trusted Platform Module (TPM). However, unlike the integrated secure engine, SE and TPM are more complex to incorporate and much more costly.

\begin{figure}[h]
  \centering
  \includegraphics[width=\linewidth]{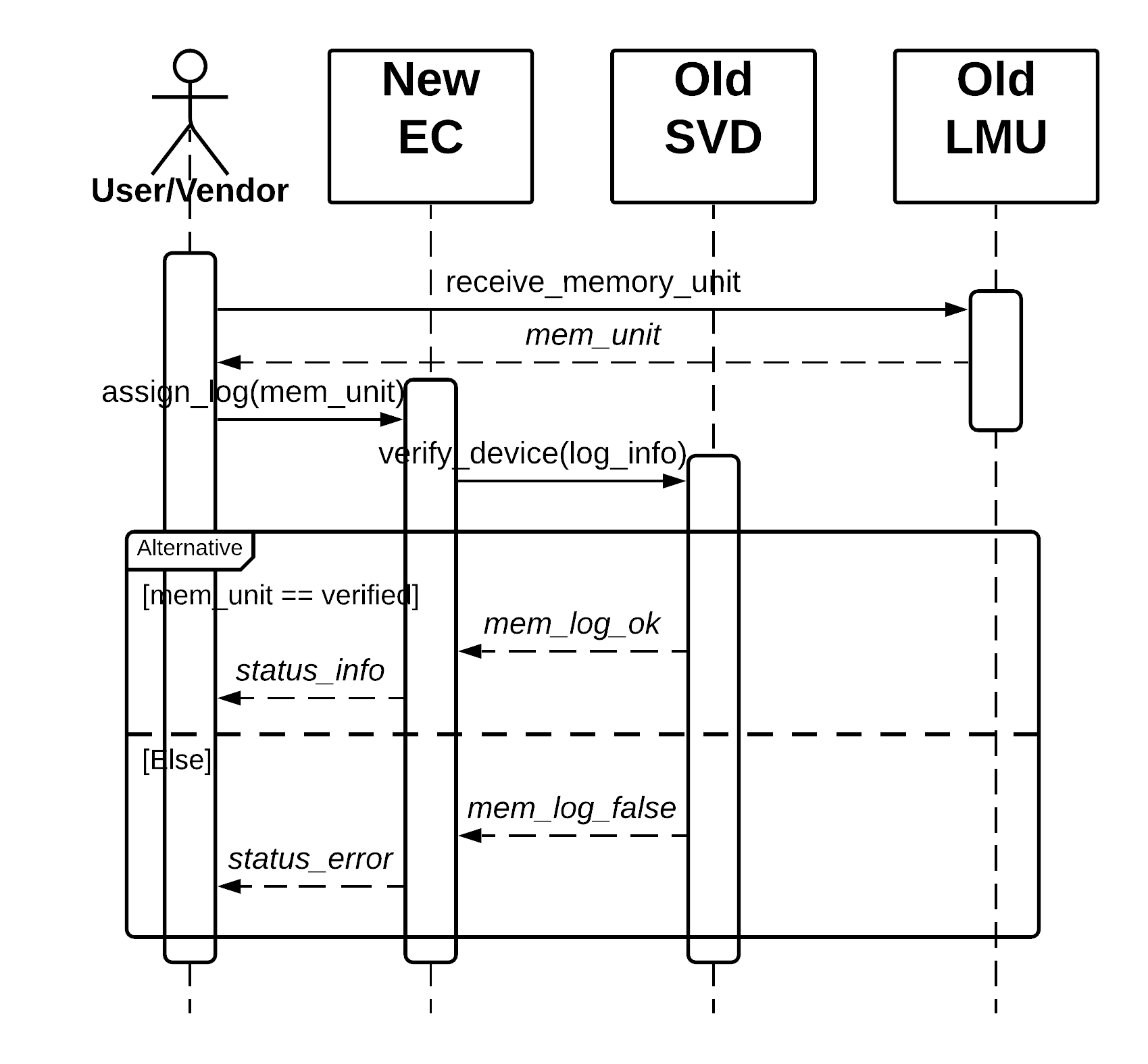}
  \caption{Sequence diagram describing the verification process during the porting of LMU and a SVD from a previous to a new embedded device platform.}
  \Description{Sequence diagram displaying the main sequence use-case when users or vendors port the logging memory unit and verify it on a new device. Contains sequence lines for the user/vendors, new device EC, Old SVD, and the Old Socket, representing the previous device.}
  \label{fig:ec2m_seq_2}
\end{figure}

The pattern is additionally aimed at providing an affordable and secure solution when transporting and then replacing an LMU. This process is depicted in Figure~\ref{fig:ec2m_seq_2}. Here, a user would receive the LMU together with the SVD from a previous socket. When integrating it into the new system, it might be necessary to verify this memory unit alongside the newly installed SVD, which has been formerly taken out from the older system. This is achieved by using the previously explained security verification functions that the new EC, through its design with SM, would possess as well. The verification process needs to be successfully completed for the LMU to be further used, be it just for the analytic or for continuing operations.

\subsection{Consequences}
This section lists the benefits and liabilities found when applying the SEL architectural pattern based on the Forces from Section~\ref{pattern_2_forces}.\\ 
The benefits of the SEL pattern are:
\begin{itemize}
    \item [F1] As the pattern suggests using a dedicated security module and predefined security functions per EC, the general production design can be applied on a larger scale.
    \item [F4] The pattern proposes the use of a dedicated SM that should allow, at a minimum, encryption and integrity check for handling the security of the stored data.
    \item [F5] Additionally, the SM needs to allow for a method of authentication and verification of individual memory modules that were previously tied to a specific pair of EC and MED. 
\end{itemize}

The liabilities when using the SEL pattern are:
\begin{itemize}
    \item [F1] As long as the security logging is only handled in a closed local embedded platform, further system updates and configurations are not handled with the proposed pattern.
    \item [F2] Each device in the suggested embedded platform is handled as a separate unit, meaning that each embedded controller comes with their own security module. This advantage at flexibility comes also with a drawback, and that is the increase of the general production cost. 
    \item [F3] Many embedded devices today are limited in terms of the extension capabilities, i.e., either not containing their own security modules or not providing additional interfaces.  
\end{itemize}

\subsection{Realized Example}
Based on the open design question presented in Section~\ref{sec:pattern_2_motivating_example}, we present a solution in form of a module extension. 
Here, security is applied to guarantee: (i) confidentiality - protecting necessary system data by only providing data associated to the BMS operational cycle, (ii) repudiation – an action can be tied to the entity that caused it, and (iii) integrity – data has not been modified.

\begin{figure}[h]
  \centering
  \includegraphics[width=\linewidth]{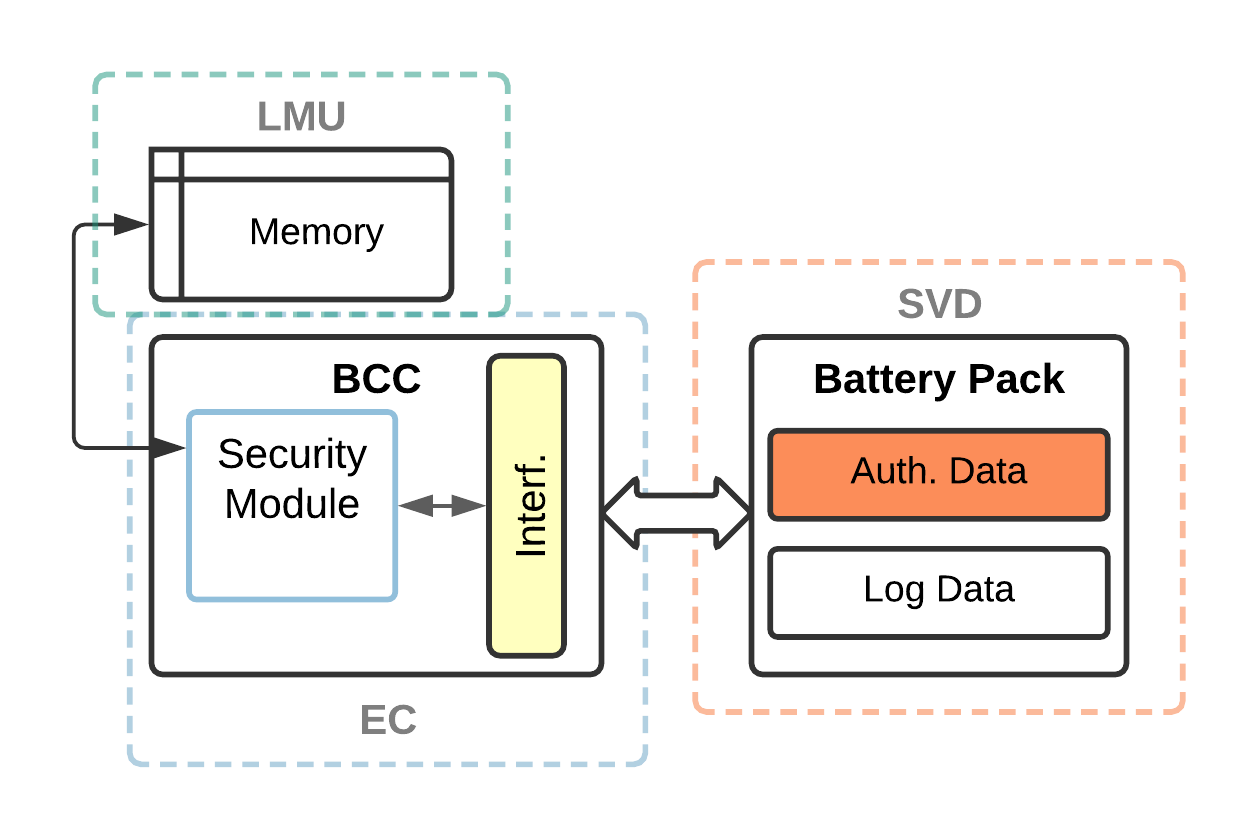}
  \caption{Realized example based on the \textsc{Secure Embedded Logging} pattern. Applied on a BCC of a BMS by extending its applicability for the secure logging process.}
  \Description{Block placement of necessary devices, showcasing the use of EC and its SM, as well as the LMU and the SVD.}
  \label{fig:sel_bms_example}
\end{figure}

The resulted block design is shown in Figure~\ref{fig:sel_bms_example}. The security functionality is controlled by an internal SM service engine. The SM communicates directly with the EC and the memory interface.

When the need arises for a battery pack to be replaced, using this design, it is possible to also easily port the whole BCC or just the memory module with the logged data, as indicated in the pattern solution. Based on the implementation, the BMS would verify the newly added BCCs, while a BCC can independently run the verification operation on the connected battery pack and memory module.

\subsection{Related Patterns}
A pattern that is similar in design but different in intent and context would be the \textsc{Security Logger and Auditor}~\cite{10.5555/2531565}. It is focused on logging security-sensitive actions from different users. Hence, this pattern offers a security solution in tying recorded information with the particular users of a system on an architectural level. Another similar pattern would be the \textsc{Secure Logger} which is traditionally used for capturing targeted application events~\cite{steel2005}. It is an implementation design pattern that can be applied on the software level in situations where otherwise system constraints are of no concern and are not taken into the design consideration.

\section{Related Work}
The patterns presented in this work are focused on delivering a design-level solution when processing data logging by providing task separation and secure handling in embedded devices. To successfully use the secure logging functions, it is necessary to implement them. This process can be done by expanding their use through one of the software-focused patterns. Several logging, and even secure logging, design patterns were already previously researched and published. Among them we have:
\begin{itemize}
    \item The \textsc{Secure Logger}~\cite{steel2005}, which is prominent, as it provides a simple solution for handling logging in different systems on an implementation level. 
    \item \textsc{Security Logger and Auditor}~\cite{10.5555/2531565} is an another pattern that provides a conceptual solution for the logging and auditing with protection mechanisms for the logged information. It deals with the repudiation aspects of information by tracking and linking the users with the logged actions.
    \item Traditionally, \textsc{Chain of Responsibility} pattern~\cite{10.5555/186897} is also often used for logging implementations as well. To make sure that the condition is met which guarantees that logging will be made in a secure manner, the secure variant of the \textsc{Chain of Responsibility} pattern can be applied~\cite{DoughertySecureDesign2009}.
\end{itemize}

During the deployment, the communication interface between MPU, EC, and MED might remain insecure. Work presented in~\cite{10.1145/3011784.3011797} presents an answer, where the \textsc{Symmetric Key Cryptography} pattern can be used for establishing a secure communication channel. Based on the system use-case, it might also be necessary to supervise and store the secured logging data online rather than locally. \textsc{Collaborative Monitoring and Logging} can be used as a template to handle the remote side of service, with an additional pattern like \textsc{Secure External Cloud Connection} used to establish the now necessary secure connection ~\cite{10.5555/2810076}. Furthermore, the work presented in ~\cite{10.1145/3147704.3147720} lists three distinct but related design patterns that can be utilized to build a remote messaging interface. These can be applied to the presented EC2M pattern to extend the abstraction on the memory unit, which in this case would mean replacing the extensible memory module with a cloud service.

\section{Conclusion}
In this paper, we have demonstrated how it is possible to implement a secure and efficient logging solution even in closed and constrained embedded systems through a careful design and separation of tasks and modules. Furthermore, necessary steps are proposed where, in case of faults or unauthorized actions, the security side would be adequately handled by giving a guide on the placement and use of a security module. This becomes increasingly important when the need for replacement and update of the active embedded devices arises, as it is necessary to also port history log data of the device’s lifetime. Lastly, the work presented in this paper is meant to encourage different vendors to consider the implementation of the secure logging functionality in local devices by not breaking the initial cost and size limitations and to help users in employing and maintaining the provided logging services for the continuing device utilization. More importantly, it streamlines the availability of the logged data to the users through simplification of transfer and security verification of memory units.

\begin{acks}
We would like to express our sincere thanks to our shepherd Jari Rauhamaki, who did a great job in providing advice and helping us to push this paper to the state that it is currently in. Furthermore, we want to give special thanks to our writer's workshop group at the EuroPLoP 2021 conference for many of their additional helpful comments and insights regarding the patterns.

This work was done within the funding project ``EFREtop: Securely Applied Machine Learning - Battery Management Systems'' (Acronym ``SEAMAL BMS'', FFG Nr. 880564).
\end{acks}

\bibliographystyle{ACM-Reference-Format}
\bibliography{paper_references}


\begin{thebibliography}{16}

%%% the Web address.

\ifx \showCODEN    \undefined \def \showCODEN     #1{\unskip}     \fi
\ifx \showDOI      \undefined \def \showDOI       #1{#1}\fi
\ifx \showISBNx    \undefined \def \showISBNx     #1{\unskip}     \fi
\ifx \showISBNxiii \undefined \def \showISBNxiii  #1{\unskip}     \fi
\ifx \showISSN     \undefined \def \showISSN      #1{\unskip}     \fi
\ifx \showLCCN     \undefined \def \showLCCN      #1{\unskip}     \fi
\ifx \shownote     \undefined \def \shownote      #1{#1}          \fi
\ifx \showarticletitle \undefined \def \showarticletitle #1{#1}   \fi
\ifx \showURL      \undefined \def \showURL       {\relax}        \fi
% invisible to TeX
\providecommand\bibfield[2]{#2}
\providecommand\bibinfo[2]{#2}
\providecommand\natexlab[1]{#1}
\providecommand\showeprint[2][]{arXiv:#2}

\bibitem[Andrea(2010)]%
        {andrea2010battery}
\bibfield{author}{\bibinfo{person}{D. Andrea}.}
  \bibinfo{year}{2010}\natexlab{}.
\newblock \bibinfo{booktitle}{\emph{Battery Management Systems for Large
  Lithium-ion Battery Packs}}.
\newblock \bibinfo{publisher}{Artech House}.
\newblock
\showISBNx{9781608071043}
\showLCCN{2012371528}
\urldef\tempurl%
\url{https://books.google.at/books?id=nivOtAEACAAJ}
\showURL{%
\tempurl}


\bibitem[David and Peterman(2014)]%
        {blackBoxesPolicy}
\bibfield{author}{\bibinfo{person}{Randall David} {and}
  \bibinfo{person}{Bill~Canis Peterman}.} \bibinfo{year}{2014}\natexlab{}.
\newblock \bibinfo{booktitle}{\emph{“Black Boxes” in Passenger Vehicles:
  Policy Issues}}.
\newblock \bibinfo{type}{Technical Report}. \bibinfo{institution}{Congressional
  Research Service}, \bibinfo{address}{Washington D.C.}
\newblock


\bibitem[Deng et~al\mbox{.}(2014)]%
        {10.1007/978-3-662-45286-8_4}
\bibfield{author}{\bibinfo{person}{Jing Deng}, \bibinfo{person}{Kang Li},
  \bibinfo{person}{David Laverty}, \bibinfo{person}{Weihua Deng}, {and}
  \bibinfo{person}{Yusheng Xue}.} \bibinfo{year}{2014}\natexlab{}.
\newblock \showarticletitle{Li-Ion Battery Management System for Electric
  Vehicles - A Practical Guide}. In \bibinfo{booktitle}{\emph{Intelligent
  Computing in Smart Grid and Electrical Vehicles}},
  \bibfield{editor}{\bibinfo{person}{Kang Li}, \bibinfo{person}{Yusheng Xue},
  \bibinfo{person}{Shumei Cui}, {and} \bibinfo{person}{Qun Niu}} (Eds.).
  \bibinfo{publisher}{Springer Berlin Heidelberg}, \bibinfo{address}{Berlin,
  Heidelberg}, \bibinfo{pages}{32--44}.
\newblock
\showISBNx{978-3-662-45286-8}


\bibitem[Dougherty et~al\mbox{.}(2009)]%
        {DoughertySecureDesign2009}
\bibfield{author}{\bibinfo{person}{Chad Dougherty}, \bibinfo{person}{Kirk
  Sayre}, \bibinfo{person}{Robert Seacord}, \bibinfo{person}{David Svoboda},
  {and} \bibinfo{person}{Kazuya Togashi}.} \bibinfo{year}{2009}\natexlab{}.
\newblock \bibinfo{booktitle}{\emph{Secure Design Patterns}}.
\newblock \bibinfo{type}{{T}echnical {R}eport} CMU/SEI-2009-TR-010.
  \bibinfo{institution}{Software Engineering Institute, Carnegie Mellon
  University}, \bibinfo{address}{Pittsburgh, PA}.
\newblock
\urldef\tempurl%
\url{https://doi.org/10.1184/R1/6583640.v1}
\showDOI{\tempurl}


\bibitem[Erl et~al\mbox{.}(2015)]%
        {10.5555/2810076}
\bibfield{author}{\bibinfo{person}{Thomas Erl}, \bibinfo{person}{Robert Cope},
  {and} \bibinfo{person}{Amin Naserpour}.} \bibinfo{year}{2015}\natexlab{}.
\newblock \bibinfo{booktitle}{\emph{Cloud Computing Design Patterns}
  (\bibinfo{edition}{1st} ed.)}.
\newblock \bibinfo{publisher}{Prentice Hall Press}, \bibinfo{address}{USA}.
\newblock
\showISBNx{0133858561}


\bibitem[Fernandez-Buglioni(2013)]%
        {10.5555/2531565}
\bibfield{author}{\bibinfo{person}{Eduardo Fernandez-Buglioni}.}
  \bibinfo{year}{2013}\natexlab{}.
\newblock \bibinfo{booktitle}{\emph{Security Patterns in Practice: Designing
  Secure Architectures Using Software Patterns} (\bibinfo{edition}{1st} ed.)}.
\newblock \bibinfo{publisher}{Wiley Publishing}.
\newblock
\showISBNx{1119998948}


\bibitem[Gamma et~al\mbox{.}(1995)]%
        {10.5555/186897}
\bibfield{author}{\bibinfo{person}{Erich Gamma}, \bibinfo{person}{Richard
  Helm}, \bibinfo{person}{Ralph Johnson}, {and} \bibinfo{person}{John
  Vlissides}.} \bibinfo{year}{1995}\natexlab{}.
\newblock \bibinfo{booktitle}{\emph{Design Patterns: Elements of Reusable
  Object-Oriented Software}}.
\newblock \bibinfo{publisher}{Addison-Wesley Longman Publishing Co., Inc.},
  \bibinfo{address}{USA}.
\newblock
\showISBNx{0201633612}


\bibitem[{Györök} and {Beszédes}(2017)]%
        {gyorok2017}
\bibfield{author}{\bibinfo{person}{György {Györök}} {and}
  \bibinfo{person}{Bertalan {Beszédes}}.} \bibinfo{year}{2017}\natexlab{}.
\newblock \showarticletitle{Fault-tolerant Software Solutions in
  Microcontroller Based Systems}. In \bibinfo{booktitle}{\emph{Orosz Gábor
  Tamás. AIS 2017–12th International Symposium on Applied Informatics and
  Related Areas: Proceedings. Székesfehérvár Magyarország 2017.11.09.
  Székesfehérvár}}. \bibinfo{pages}{7--12}.
\newblock
\showISBNx{978-963-449-032-6}


\bibitem[{Györök} and {Beszédes}(2018)]%
        {gyorok2018}
\bibfield{author}{\bibinfo{person}{György {Györök}} {and}
  \bibinfo{person}{Bertalan {Beszédes}}.} \bibinfo{year}{2018}\natexlab{}.
\newblock \showarticletitle{Highly reliable data logging in embedded systems}.
  In \bibinfo{booktitle}{\emph{2018 IEEE 16th World Symposium on Applied
  Machine Intelligence and Informatics (SAMI)}}.
  \bibinfo{pages}{000049--000054}.
\newblock
\urldef\tempurl%
\url{https://doi.org/10.1109/SAMI.2018.8323985}
\showDOI{\tempurl}


\bibitem[Inc.(2020)]%
        {videoSurveillanceQumulo}
\bibfield{author}{\bibinfo{person}{Qumulo Inc.}}
  \bibinfo{year}{2020}\natexlab{}.
\newblock \bibinfo{booktitle}{\emph{Video Surveillance File Data Solutions}}.
\newblock
\urldef\tempurl%
\url{https://qumulo.com/solution/surveillance/}
\showURL{%
Retrieved Jan 03, 2021 from \tempurl}


\bibitem[LLC(2020)]%
        {videoSurveillanceSeagate}
\bibfield{author}{\bibinfo{person}{Seagate~Technology LLC}.}
  \bibinfo{year}{2020}\natexlab{}.
\newblock \bibinfo{booktitle}{\emph{Video Surveillance Storage: How Much Is
  Enough?}}
\newblock
\urldef\tempurl%
\url{https://www.seagate.com/gb/en/solutions/surveillance/how-much-video-surveillance-storage-is-enough/}
\showURL{%
Retrieved Jan 03, 2021 from \tempurl}


\bibitem[Mobility and Transport(2021)]%
        {blackBoxesEU}
\bibfield{author}{\bibinfo{person}{European~Commission: Mobility} {and}
  \bibinfo{person}{Transport}.} \bibinfo{year}{2021}\natexlab{}.
\newblock \bibinfo{booktitle}{\emph{Black boxes/ in-vehicle data recorders}}.
\newblock
\urldef\tempurl%
\url{https://ec.europa.eu/transport/road_safety/specialist/knowledge/esave/esafety_measures_known_safety_effects/black_boxes_in_vehicle_data_recorders_en}
\showURL{%
Retrieved Jan 03, 2021 from \tempurl}


\bibitem[Sinnhofer et~al\mbox{.}(2016)]%
        {10.1145/3011784.3011797}
\bibfield{author}{\bibinfo{person}{Andreas~Daniel Sinnhofer},
  \bibinfo{person}{Felix~Jonathan Oppermann}, \bibinfo{person}{Klaus
  Potzmader}, \bibinfo{person}{Clemens Orthacker}, \bibinfo{person}{Christian
  Steger}, {and} \bibinfo{person}{Christian Kreiner}.}
  \bibinfo{year}{2016}\natexlab{}.
\newblock \showarticletitle{Patterns to Establish a Secure Communication
  Channel}. In \bibinfo{booktitle}{\emph{Proceedings of the 21st European
  Conference on Pattern Languages of Programs}} (Kaufbeuren, Germany)
  \emph{(\bibinfo{series}{EuroPlop '16})}. \bibinfo{publisher}{Association for
  Computing Machinery}, \bibinfo{address}{New York, NY, USA}, Article
  \bibinfo{articleno}{13}, \bibinfo{numpages}{21}~pages.
\newblock
\showISBNx{9781450340748}
\urldef\tempurl%
\url{https://doi.org/10.1145/3011784.3011797}
\showDOI{\tempurl}


\bibitem[Sousa et~al\mbox{.}(2017)]%
        {10.1145/3147704.3147720}
\bibfield{author}{\bibinfo{person}{Tiago~Boldt Sousa},
  \bibinfo{person}{Hugo~Sereno Ferreira}, \bibinfo{person}{Filipe~Figueiredo
  Correia}, {and} \bibinfo{person}{Ademar Aguiar}.}
  \bibinfo{year}{2017}\natexlab{}.
\newblock \showarticletitle{Engineering Software for the Cloud: Messaging
  Systems and Logging}. In \bibinfo{booktitle}{\emph{Proceedings of the 22nd
  European Conference on Pattern Languages of Programs}} (Irsee, Germany)
  \emph{(\bibinfo{series}{EuroPLoP '17})}. \bibinfo{publisher}{Association for
  Computing Machinery}, \bibinfo{address}{New York, NY, USA}, Article
  \bibinfo{articleno}{14}, \bibinfo{numpages}{14}~pages.
\newblock
\showISBNx{9781450348485}
\urldef\tempurl%
\url{https://doi.org/10.1145/3147704.3147720}
\showDOI{\tempurl}


\bibitem[Steel et~al\mbox{.}(2005)]%
        {steel2005}
\bibfield{author}{\bibinfo{person}{Christopher Steel}, \bibinfo{person}{Ramesh
  Nagappan}, {and} \bibinfo{person}{Ray Lai}.} \bibinfo{year}{2005}\natexlab{}.
\newblock \bibinfo{booktitle}{\emph{Core Security Patterns: Best Practices and
  Strategies for J2EE, Web Services, and Identity Management}}.
\newblock \bibinfo{publisher}{Pearson}, \bibinfo{address}{USA}.
\newblock
\showISBNx{0131463071}


\bibitem[Xiong et~al\mbox{.}(2018)]%
        {8168251}
\bibfield{author}{\bibinfo{person}{Rui Xiong}, \bibinfo{person}{Jiayi Cao},
  \bibinfo{person}{Quanqing Yu}, \bibinfo{person}{Hongwen He}, {and}
  \bibinfo{person}{Fengchun Sun}.} \bibinfo{year}{2018}\natexlab{}.
\newblock \showarticletitle{Critical Review on the Battery State of Charge
  Estimation Methods for Electric Vehicles}.
\newblock \bibinfo{journal}{\emph{IEEE Access}}  \bibinfo{volume}{6}
  (\bibinfo{year}{2018}), \bibinfo{pages}{1832--1843}.
\newblock
\urldef\tempurl%
\url{https://doi.org/10.1109/ACCESS.2017.2780258}
\showDOI{\tempurl}


\end{thebibliography}

% \appendix

\end{document}